\documentclass[10pt]{iopart}

\usepackage{graphicx}
\usepackage{xcolor}
\usepackage{float}
\usepackage[utf8]{inputenc}
\usepackage[T1]{fontenc}

\begin{document}
	
\title[Tunable frequency dependent stiffness elements]{Implementation of tunable frequency-dependent stiffness elements via integrated shunted piezoelectric stacks}
	
\author{B Van Damme$^1$, R Weber$^2$, J U Schmied$^3$, A Spierings$^2$ and A Bergamini$^1$}
 
\address{$^1$ Empa, Materials Science and Technology, Laboratory for Acoustics/Noise Control, CH-8600, Dübendorf.}	
\address{$^2$ Inspire AG, Innovation centre for additive manufacturing.}
\address{$^3$ ETH Zürich, Laboratory of Composite Materials and Adaptive Structures}
\ead{bart.vandamme@empa.ch}	

\begin{abstract}
Piezoelectric transducers applied on or integrated in structures, combined with appropriate circuits have been extensively investigated as a smart approach to the mitigation of resonant vibrations with high relative amplitudes. A resonant shunt circuit consisting of the capacitive piezoelectric transducer and an inductance can be configured to target specific eigenmodes of a structure, if appropriately placed and tuned. Their effect is expressed in terms of mechanical impedance of the host structure, allowing for the exchange of energy between the mechanical and electrical domain, to dramatically affect the dynamic response of the structure. By re-framing the function of resonant shunted piezoelectric transducers as frequency dependent variable stiffness elements, this paper investigates their capability to realize a frequency dependent structural mechanical connectivity, where the load path within a lattice structure can be interrupted at will for specific frequencies by tunable null-stiffness components. Here, we offer the numerical and experimental verification of this idea,  by demonstrating the ability to significantly affect the dynamic response of a unit cell of an adaptive lattice metamaterial, even away from a structural resonance. In the latter case, the null-stiffness shunt leads to an additional resonance peak in the truss' dynamic response. Its realization as additively manufactured component points to the feasibility of such structures in real life.  
\end{abstract}

\submitto{\SMS}
\maketitle

\section{\label{sec:Intro}Introduction}
\begin{figure} [ht]
    \centering
    \includegraphics{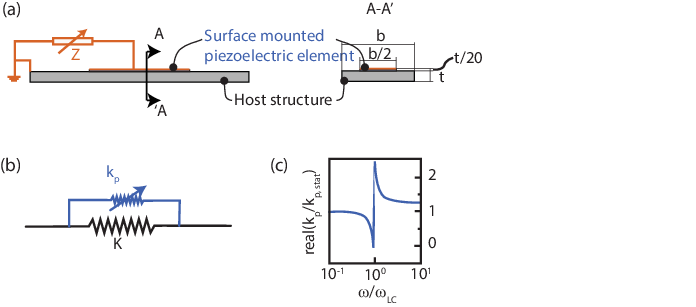}
    \caption{a) Surface mounted piezoelectric element, as a typical example of integration strategy for vibration control properties b) representation of the host structure with surface mounted piezoelectric element (as described in \cite{Delpero2012}) as two spring elements $K$ and $k_p$ in  parallel, with the piezoelectric as variable stiffness element associated with the variable impedance $Z$ of the electric shunt. c) stiffness change (real part) around the electrical resonance $\omega_{e}$ of a resonantly (Z=L) shunted transducer, adapted from \cite{Delpero2012_diss, Hagood1991}.}
    \label{Fig:parallel}
\end{figure}

 The implementation of linearly \cite{vanSpregen2022,Lossouarn2024,Berardengo2020,Chatziathanasiou2022} or non-linearly \cite{Alfahmi2024} shunted piezoelectric elements in smart structures for vibration reduction is often associated with the surface application \cite{Zhou2023,Mazur2021} of transducers onto plate like structures. An inductive shunt in series with a  piezo transducer results in an electrical circuit resonance, which interacts with the dynamics of the host structure. This set-up is very well suited for plate-like structures. In correspondence with the large vibrational deformation at a resonant frequency $\omega_0$ of the host structure with bending stiffness $K$, the coupling between the shunted transducer and the host structure (Fig.~\ref{Fig:parallel}) is sufficient to significantly affect the dynamic response of the whole structure in proximity of resonance, as reported by numerous sources \cite{Hagood1991, Saravanos1999DAMPEDELEMENTS, Fleming2003, Niederberger2004AdaptiveDamping, Porfiri2007, Delpero2012, Lossouarn2016MultimodalNetwork,Toftekr2020ExperimentalStructures}. However, the ratio of static stiffness between transducer and host structure is usually quite small. For example, in the cases discussed in \cite{Delpero2012} and exemplified in Fig.~\ref{Fig:parallel} a), the addition of the transducer increases the bending stiffness of the plate by about 9\%. The stiffness enhancement stems from the parallel arrangement of $k_p$ and $K$ shown in Fig.~\ref{Fig:parallel} b) can be calculated from the increase in second moment of area of the bending element, as the bending modes typically correspond to the vibration of interest in this kind of application. Here, we assume that the elastic modulus of the piezoelectric material and the host structure are the same, a good approximation for the case of Lead Zirconate Titanate (PZT) and Aluminum ($E_{PZT}\approx E_{Al}=70GPa$).  However, the dynamic stiffness $k_p$ of the piezoelectric material is well known to depend on the electrical boundary conditions applied to the transducer. In the case of a resonant shunt, it changes with frequency as shown in \ref{Fig:parallel} c)  and reported in~\cite{Hagood1991,Delpero2012}, with an asymptotic stiffness variation around the $LC$-shunt resonance. Away from the mechanical resonant condition, the variation of the piezoelectric stiffness $k_p$, as shown in Fig.~\ref{Fig:parallel} c), is hardly discernible in the dynamic response of the structure. Therefore, the parallel structure-piezo arrangement only shows its effect on the dynamic response when the electrical and mechanical resonance frequencies coincide.
This parallel arrangement of transducer and plate-like host structure offers significant advantages in terms of structural integrity compared to the series arrangement discussed later, as the strain experienced by the brittle piezoelectric material it limited by the typically very stiff host structure. 
    
\begin{figure}[ht]
    \centering
    \includegraphics{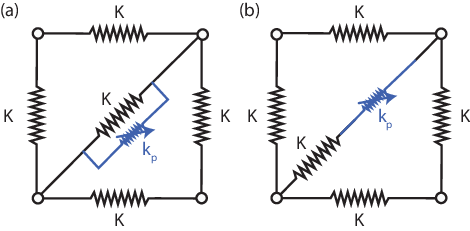}
    \caption{Piezoelectric element as variable stiffness ($k_p$) element surface mounted on the diagonal brace (left) or inserted in the brace (right) of a truss structure made of beams of arbitrary stiffness $K$.}
    \label{Fig:brace}
\end{figure}

When the deformation of structural elements is dominated by the axial component, as in the struts of a truss structure, the limited ability of a parallel arrangement of surface mounted elements to change the (bending) stiffness of a beam or plate, as shown in Fig \ref{Fig:brace} a) becomes even more apparent: assuming the same ratio of cross-sectional areas shown Fig.~\ref{Fig:parallel} a), the contribution of the surface-mounted piezoelectric element to the axial stiffness becomes:
\begin{equation}
 K_{tot}=K+k_p=E_{Al,PZT}(A_{Al}+A_{p}),
 \label{Eq:parallel}
\end{equation}
where $A_{Al}$ and $A_{p}$ are the cross-sectional areas of the Aluminum strut and the piezo, respectively. With the area values shown in Fig.~\ref{Fig:parallel} a), the contribution of the piezo to the strut stiffness is further reduced to 2.5\%. Here, the strong contribution of the off-axis stiffness of the surface mounted element, proportional to the square of the distance to the neutral axis, seen in bending deformations, is lost under axial loads. 

The dynamic responses of the systems at hand are determined by the mass and stiffness matrices. The host-structure dominated stiffness as in the case of a parallel arrangement can only be significantly affected by the frequency dependent dynamic stiffness of the piezoelectric transducer in correspondence of the stiffness minima occurring at resonance, given  sufficient generalized electro-mechanical coupling.

 The integration of piezoelectric elements in a series arrangement relative to the local load path as an approach to the reduction of resonant vibrations has also been reported in literature \cite{Hagood1991b,Preumont2008,Mayer2016,Weber2022IntegrationFusion}, mainly because they offer a high electromechanical coupling coefficient of the transducer in the struts constituting truss structures.

 In the presented embodiment of this kind of integration, one or a finite number of piezoelectric elements are integrated as a series element into individual strut(s) and in parallel with other struts in the same section of the global load path. Within an individual strut, the stiffness contribution of the piezoelectric element can be written as:
\begin{equation}
\frac{1}{K_{tot}}=\frac{1}{K}+\frac{1}{k_p}
\label{Eq:series}
\end{equation}
As the comparison of the contribution of the piezoelectric transducer described by equations \ref{Eq:parallel} and \ref{Eq:series} shows, the stiffness in the parallel arrangement is prevalently determined by the stiffer component, while in the series arrangement the more compliant element is most significant. 

Similar to the classical parallel arrangement for bending modes, the shunted piezoelectric resonator is coupled with a specific resonant frequency of the structure as a means to extract mechanical energy, as an electro-mechanical equivalent of a tuned mass damper. Like in the case of the surface mounted transducers, appropriate tuning of the piezoelectric resonator to the global mechanical mode that needs to be addressed is key to the emergence of the function of the transducer. However, in the series arrangement, the resonant transducer is now an integrated part of the vibrating structure.

In a different interpretation of the functionality of linearly shunted piezoelectric elements, previous work \cite{Bergamini2014}, has shown an implementation of null-stiffness elements in phononic crystals. There, the scattering centers of a phononic crystal inspired by the work of Wu \cite{Wu2008} could be effectively disconnected from the wave guide, using null stiffness elements consisting of the resonantly shunted piezoelectric discs interposed between the scatterers and a continuous medium. The result of this was disappearance of a phononic bandgap in a frequency range selected by tuning the resonance of a series of LC circuits. In this work, we move one step forward towards structures with adaptive dynamics, and numerically and experimentally demonstrate frequency dependent connectivity switching using a piezoelectric stack in a braced metallic frame, also in a non-resonant regime. The 3D-printed aluminium frame with integrated piezo stack was originally reported in \cite{Weber2022IntegrationFusion}. The out of resonance stiffness modification of a unit cell of an additively manufactured lattice material can be seen as a step towards the implementation of adaptive metamaterials postulated in \cite{Schmied2019}.

\begin{figure}[ht]
    \begin{center}
        \includegraphics{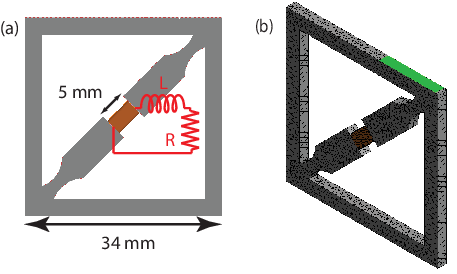}
    \end{center}
    \caption{(a) The unit cell structure with integrated piezoelectric stack element, inductively shunted to achieve frequency dependent stiffness. (b) Mesh of the unit cell generated in Ansys for the numerical analysis. The green area shows the constrained surface of the structure.}
    \label{Fig:structure}
\end{figure}

In the following, we first present the concept of variable stiffness elements by showing their potential in finite element models. Then, we present an experimental study using variable synthetic inductive shunts based on Antoniou-circuits~\cite{antoniou1969}. The conclusion describes the results of shunts in the resonant and non-resonant frequency ranges.

\section{Finite element modeling of the variable stiffness element}\label{sec:Model}
\subsection{Variable stiffness of the piezo stack with shunt}
The interaction between an   inductively shunted piezo patch and the host structure can be explained in function of a change in mechanical impedance, as described by~\cite{Hagood1991}. The impedance ratio between the shunted and open-circuit piezo patch is given by the formula 
\begin{equation}\label{eq:impedance}
    Z = 1-k^2_{33}\left(\frac{\delta^2}{\gamma^2+\delta^2r\gamma+\delta^2}\right),
\end{equation}
where $\delta=\omega_e/\omega_n$ is the ratio between the electrical shunt frequency $\omega_e=1/\sqrt{LC}$ and the mechanical natural frequency of the piezo patch. The normalized frequency variable is $\gamma = i\omega /\omega_n$, and $r=RC\omega_n$ describes the damping due to the resistance in the shunt. Finally, 
\begin{equation}
    k_{33}=\frac{d_{33}}{\sqrt{s_{33}\varepsilon_3}}
\end{equation}
includes the electromechanical coupling properties of the piezo patch, with $d_{33}$ being the piezoelectric coupling coefficient (in C/N), $s_{33}$ the mechanical compliance (the inverse of the Young's modulus), and $\varepsilon_3$ the electrostatic permittivity, in the corresponding directions.

\begin{figure}
    \centering
    \includegraphics{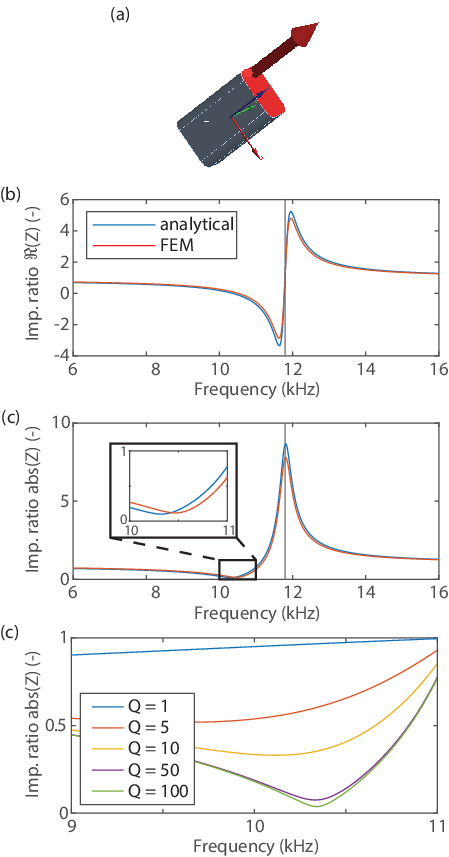}
    \caption{Tunable mechanical impedance of the piezo element. (a) Finite element setup, showing the piezo orientation along the blue $z$-axis, and the applied force on the top electrode in red. The bottom electrode is blocked. (b) Real part of the mechanical impedance ratio according to eq.~(\ref{eq:impedance}) and calculated using the finite element model. The chosen inductance is $L=10$~H and the resistance $R=20$~k$\Omega$. (c) Absolute value of the mechanical impedance ratio shown in (b). The zoomed section shows a minimum at a ratio 0.12 compared to the static stiffness. (d) Analytical calculation of the absolute value of the impedance ratio, showing that the stiffness reduction is less pronounced for values of the Q-factor below 50.}
    \label{fig:impedance}
\end{figure}

For the case described in this paper, where a piezo material is integrated in the mechanical structure to change its overall dynamic stiffness, a high reduction of the mechanical impedance is required. At first, it is verified that a finite element model of the shunted piezoelectric patch behaves as predicted by theory. The model is constructed in Ansys 2020R2. Its properties are given in Tab.~\ref{tab:PZT}, and are implemented using Ansys' Piezoelectric and MEMS toolbox, which also allows to define the top and bottom of the piezo element as electrodes by coupling the voltage of all finite element nodes on these surfaces. The capacitance of the piezo part, as retrieved from the model, is 22.7~pF. This is in line with the expected value for a monolithic piezo crystal with these dimensions. The inductive shunt is implemented using an APDL script. A single node is newly defined, and two CIRCU94 elements are defined between this node and the two electrodes of the PZT element. One element represents the shunt, the other a resistor which allows to change the Q factor of the shunt. For two cases, open circuit and shunted, a harmonic unit force is applied on the top electrode while the bottom side of the piezo is rigidly blocked (Fig.~\ref{fig:impedance} (a)). The complex frequency response function of the displacement of the top electrode is extracted between 6 and 16~kHz, so that the mechanical complex stiffness can be calculated.

\begin{table}
    \centering
    \caption{Properties of the piezoelectric element in the model}
    \label{tab:PZT}
    \begin{tabular}{l|c}
        Density &  7800 kg/m$^3$\\
        Young's modulus & 70 GPa\\
        $d_{31}$ & -5.3 C/m$^2$\\
        $d_{33}$ & 15.8 C/m$^2$\\
        $\varepsilon_r$ & 1700
    \end{tabular}    
\end{table}

The real part and absolute value of the dynamic stiffness are shown in Fig.~\ref{fig:impedance} (b) and (c). An open-circuit simulation shows that the first longitudinal mode of the piezo material occurs at $\omega_n=200$~kHz, which is far higher than the desired inductive shunt eigenfrequencies. A characteristic decrease followed by an increase of the mechanical impedance can be noticed, with only a small difference between the analytical and numerical response, as shown in Fig.~\ref{eq:impedance} (b). This difference is due to the orthotropic mechanical and piezoelectric properties included in the finite element model, which are not captured in the analytical model. Fig.~\ref{eq:impedance} (d) illustrates the mechanical stiffness reduction as a function of the shunt's Q-factor. For values higher than 50, a more than 90\% reduction can be seen, whereas the stiffness change decreases rapidly with decreasing Q-values.

\subsection{Truss structure with variable stiffness element}
A cell of a truss made of aluminium is modelled in Ansys 2020R2. The detailed geometry of the truss, shown in Fig.~\ref{Fig:structure} (b),  represents the 3D-printed experimental specimen described in Sec.~\ref{sec:Experiment}. The diagonal strut, which defines the high rigidity for in-plane deformation, has an embedded PZT cell of 5 mm length. 

A harmonic analysis is performed on the cell without an inductive shunt: the top right corner is elastically supported, and the entire structure is excited by a 10~m/s$^2$ acceleration to mimic a body force. This reflects the experimental conditions, where the light structure is entirely excited by an electrodynamic shaker. The truss cell has 3 in-plane vibrational modes in the frequency range from 5 to 13~kHz, at roughly 6, 9.5, and 11.5~kHz. Other vibration modes, i.e. the modes with deformation shapes normal to the truss plane, are not considered for this study since the diagonal strut does not play a crucial role. 

First, we investigate the efficiency of the embedded piezo for use as a tuned vibration damper. The corresponding inductances to address the three first eigenfrequencies, gathered from the formula 
\begin{equation}
    L = \frac{1}{\omega_e^2 C}
\end{equation}
are thus 28.0~H, 11.8~H and 8.4~H. The Q-factor of the shunt can be regulated by the series resistor through 
\begin{equation}
    Q = \frac{\omega_e L}{R},
\end{equation}
and is set to Q=50 for all cases. 

\begin{figure}[ht]
    \centering
    \includegraphics{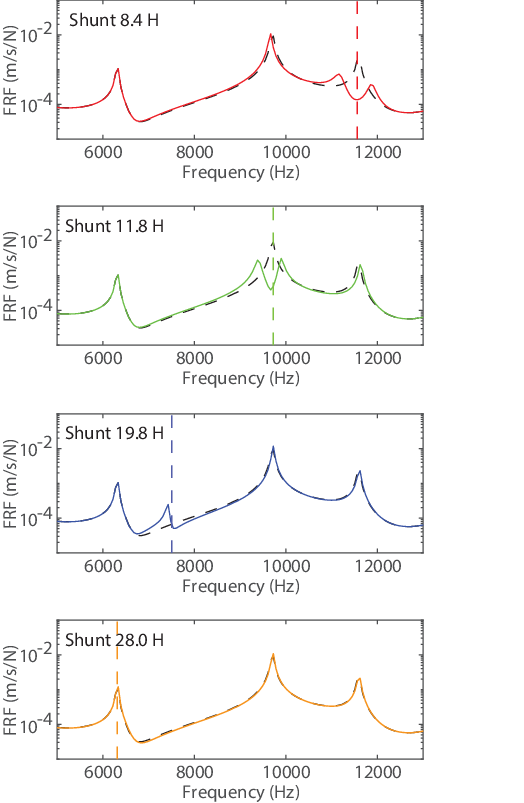}
    \caption{Simulated frequency response function of the bracket for 4 different inductive shunts. The black dashed line shows the unshunted case. The vertical dashed lines show the electrical shunt resonance frequency. The shunt tuned to the first eigenmode does not have a measurable effect, whereas correct tuning reduces the modal amplitude of the second and third peak. Mistuning the shunt leads to an additional mechanical resonance.}
    \label{fig:L_simulation}
\end{figure}

\begin{figure}[ht]
    \centering
    \includegraphics{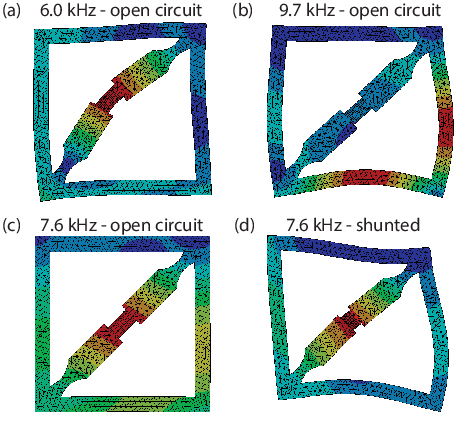}
    \caption{Modal deformation of the truss at 6.0~kHz (a) and 9.7~kHz (b). The first mode is characterized by bending of the diagonal strut, the second mode by a deformation along the axis of the strut. The color scale shows normalized vibration velocities. Deformed shape at 7.6~kHz with identical scaling factor $\times 10^6$ for the visualization of the deformation, in the open-circuit (c) and shunted (d) case. The overall deformation of the truss, and the extreme compression of the piezo due to a loss of dynamic stiffness are clearly noticeable.}
    \label{fig:modeshape}
\end{figure}

The results are shown in Fig.~\ref{fig:L_simulation}. The embedded piezo is extremely effective to attenuate the second and third mode, resulting in a pronounced reduction of the modal peak. However, the first mode is hardly affected by the shunted piezo element. The reason for this is the bending state of the diagonal strut at the first eigenfrequency (Fig.~\ref{fig:modeshape} (a)), leading to a small net charge accumulation since positive and negative charges balance each other out, and the electromechanical coupling is minimal. The piezo transducer under bending is compressed on one side and extended on the other, which inhibits large charge motion through the shunt. The second mode is shown in Fig.~\ref{fig:modeshape} (b), revealing a mainly axial deformation of the piezo element, and the strain state leads to a large charge accumulation on the electrodes of the piezo element.

The potential of the piezo stack to act as a tunable stiffness element is shown by detuning the shunt to an inductance value leading to an electrical resonance which does not coincide with a mechanical resonance. This is conventionally not of much interest, since low vibration amplitudes lead to small changes in the overall stiffness, especially in the parallel transducer arrangement. However, given the axial stress in the diagonal strut and the resonant behaviour of the shunt, this regime for the series arrangement has a measurable effect. As illustrated in Fig.~\ref{fig:L_simulation}, a shunt of 19.8~H induces a new mechanical resonance at 7.6~kHz. The piezo element loses its mechanical stiffness at the resonance frequency, thereby compromising the rigidity of the entire diagonal strut. The resulting deformation is visualized in~\ref{fig:modeshape}~(d): the piezo element loses its dynamic stiffness, thereby allowing large axial deformations of the diagonal strut, and a large overall deformation of the truss structure.

\section{Experimental validation of the variable stiffness structure}\label{sec:Experiment}
The concepts introduced in the previous section are brought to reality using a sample that was produced by rapid prototyping~\cite{Weber2022IntegrationFusion}. The truss is created using selective laser melting of aluminium powder. A piezo stack (Thorlabs AE0203D04DF) with capacitance 98.9~nF is clamped into the diagonal strut after printing, thereby ensuring mechanical contact during vibration due to a sufficiently high static precompression force. The stack's capacitance is higher than the simulated value of the piezo monocrystal in the previous section, which allows us to use inductors with values between 1 and 10~mH to cover the desired frequency range.
 
\begin{figure}[ht]
     \centering
     \includegraphics{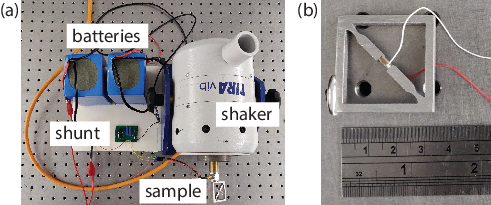}
     \caption{Experimental setup showing the Antoniou shunt circuit on the left, and the shaker with truss on the right (a). The sample with embedded piezo stack (b).}
     \label{fig:setup}
\end{figure}
 
The experimental setup is shown in Fig.~\ref{fig:setup}. The sample is mounted on an electrodynamic shaker by means of a force sensor (PCB208C01), whose output is used as reference signal for the FRF calculations. The response of the truss is measured on the vertical strut opposite the excitation point (at the bottom of the figure) by a scanning laser vibrometer (Polytec PSV-400). The excitation signal is a sine sweep ranging from 4 to 13~kHz. This spans the first three in-plane resonance frequencies. Higher frequencies could not be achieved by the shaker system, and at lower frequencies the combined system of the shaker and sample yields rigid body modes of the sample. The latter are not relevant to the study at hand, since the struts are not deformed and no effect of the piezo stack can be expected. For practical reasons, the inductance is realized by an Antoniou synthetic gyrator circuit~\cite{antoniou1969}, powered by 2 batteries. This classical electrical circuit using two operational amplifiers can be used as a tunable inductor, by varying the resistance value of a potentiometer. Apart from being able to generate inductances in the desired range, the $Q$-factor can be tuned by an additional potentiometer in series. It can be kept at a sufficiently high value to achieve a measurable variable stiffness.  
	
\begin{figure}[ht]	
    \begin{center}
        \includegraphics{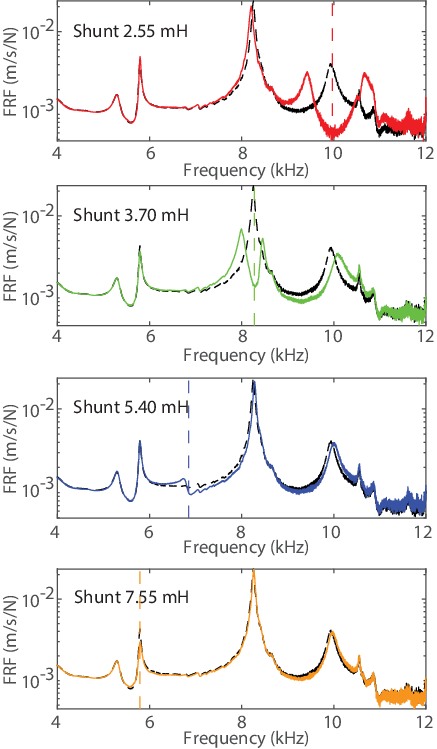}
    \end{center}
    \caption{Measured frequency response function, averaged over a vertical strut, of the bracket for 4 different inductive shunts. The black dashed line shows the unshunted case. The vertical dashed lines show the electrical shunt resonance frequencies $\omega_e=1/\sqrt{LC}$. The shunt tuned to the first eigenmode does not have a measurable effect, whereas correct tuning reduces the modal amplitude of the second and third peak. Mistuning the shunt leads to an additional mechanical resonance.}
    \label{Fig:L_results}
\end{figure}

The experimental results shown in Fig.~\ref{Fig:L_results} are in line with the simulations, showing three main resonance peaks in the chosen frequency range, and similar velocity amplitudes. The resonance frequencies show some deviation. This is due to geometrical differences, and the unknown stiffness and damping of the connection to the shaker, leading to complex boundary conditions that can only be approximately reproduced in the model. Moreover, the experimental spectrum is richer than the modeled spectrum showing several minor peaks because of the inevitable out-of-plane vibrations of the structure. For correctly tuned inductance values, the second and third eigenmodes are efficiently suppressed, whereas the first eigenmode is not affected. For an inductance leading to an electrical resonance that does not coincide with a mechanical eigenfrequency, a clear but rather feeble additional mechanical resonance can be noticed. The $Q$-factor of the Antoniou circuit is too low to achieve more pronounced effects. It is remarkable, however, that even far away from mechanical resonances the electromechanical coupling between the structure and the embedded piezo can change the dynamic stiffness.

\begin{figure}[ht]
    \centering
        \includegraphics{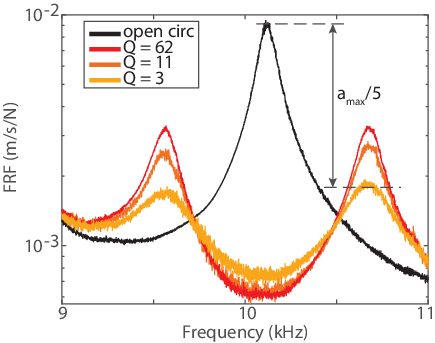}
    \caption{Shunted tuned damping of the third vibration mode (shown in black) with decreasing $Q$-factor, achievable within the limits of the Antoniou circuit. A well chosen resonance can flatten the secondary vibration peaks.}
    \label{fig:Qfactor}
\end{figure}

As is well known, changing the $Q$-factor of the shunt can have additional advantageous effects on the modal damping. This is illustrated in Fig,~\ref{fig:Qfactor} for the damping of the third eigenmode. 

\section{\label{sec:Results}Results and discussion}
A series-configuration of a piezoelectric stack built into the diagonal connection of a truss structure becomes an integral part of the static and dynamic load path of the system. Since piezoelectric materials have a Young's modulus similar to that of metals, this inclusion does not deteriorate the static stiffness of the truss. However, this configuration allows us to insert a variable-stiffness element into a load-bearing strut. We could experimentally verify the findings that a series-arrangement of a piezo stack transducer within a load-bearing element functions as a highly effective tuned damper presented in \cite{Hagood1991b} for the case of an additively manufactured structure, with a drop-in integrated piezoelectric transducer. 

More importantly, finite element simulation results show that, for inductive shunt circuits with a reasonable Q-factor, the dynamic stiffness of a piezo transducer can however be lowered sufficiently to induce a new vibration mode of the entire structure. In the case at hand, with a frequency range of interest between 5 and 15~kHz, inductance simulation circuits are able to achieve this effect in reality.

Although the quality of the circuit and its individual components did not allow to achieve a sufficiently high Q factor to achieve high-amplitude vibrations, a change in stiffness of the truss leads to higher dynamic deformation at a chosen frequency between two resonance frequencies. The electric resonance effectively removes the rigidity of the central diagonal strut, thus changing the 'mechanical topology' of the structure. Models show that a new vibration mode with an amplitude similar to the purely mechanical modes of the structure is possible. However, if the shunt's resistance is too high due to the inherent losses in the operational amplifiers, the stiffness drop and the resulting vibrational motion are less pronounced.
 
\section{\label{sec:Conclusions}Conclusions and outlook}
The experiments and the corresponding numerical models confirm that the increased electromechanical coupling, associated with a series arrangement of the piezo element within a vibrating structure, allows for a modification of the dynamic response that goes beyond the modal vibration reduction  demonstrated in \cite{Hagood1991b,Preumont2008,Mayer2016}. The excellent coupling allows for a full loss of stiffness in the diagonal strut of a truss structure, as demonstrated by the introduction of additional vibration modes stemming from the modification of the stiffness of the diagonal strut. However, the embedded piezoelectric stack is only effective if it undergoes a net strain along its polarized axis. A pure bending deformation, as is seen in the first vibration mode of the truss, does not lead to a measurable effect on the dynamics of the system.

For future applications, the size and load bearing capacity of the piezoelectric stacks will be defining factors. Although larger stacks can undergo larger deformations, have a higher capacitance, and should therefore be able to affect lower frequencies for reasonable inductance values, they are not commercially available in sizes larger than several tens of millimeters. Moreover, to achieve tunable stiffness of structures with a more complex, potentially three-dimensional, shape, the placement of the stack has to be well considered. If the local strain is not sufficiently high, the effect on the structure's dynamics will be negligible.

The tunability of the system by using an Antoniou-gyrator inductance gives full control over the vibrational properties of the system. At the level of the individual unit shown in Fig.~\ref{Fig:structure}, the loss of stiffness can for instance be beneficial to avoid damage at high loads by letting the structure yield, or to create vibrating elements that can be used as tuned vibration absorbers. From the perspective of the synthesis of novel architected materials that transcend the example discussed in \cite{Bergamini2014}, the ability to modify at will the connectivity of a lattice structures is a very appealing way to create lattice structures with frequency dependent unit cell geometry. Here, the  factor limiting the implementation of this tunable unit cell geometry scheme over a wide range of frequencies, is the range of stable inductances and fairly low $Q$-factors that can customarily be achieved by Antoniou circuit implementation of synthetic gyrators. Alternative synthetic gyrator designs may offer relief to this constraint.

\section*{Acknowledgments}
Initial research leading related to this work was supported by the Air Force Office of Scientific Research grant FA9550-15-1-0397 ``Integrated multi-field resonant metamaterials for extreme, low frequency damping.” The simulations and experiments were performed under SNSF grant CRSK-2\_190732 ``Shunted electro-active membrane absorbers".

\end{document}